\documentclass[twocolumn,pra,aps,superscriptaddress,floatfix]{revtex4}
\usepackage[latin1]{inputenc}
\usepackage{bm}
\usepackage{amssymb,amsbsy,amsmath}
\usepackage{graphicx}
\usepackage{verbatim}
\usepackage{color}
\usepackage{pifont}
\usepackage{cancel}

\begin{document}

\title{Detection of quantum non-Markovianity close to the Born-Markov
approximation}
\author{Thais de Lima Silva}
\affiliation{Instituto de Física, Universidade Federal do Rio de Janeiro, Caixa Postal
68528, Rio de Janeiro, RJ 21941-972, Brazil}

\author{Stephen P. Walborn}
\affiliation{Instituto de Física, Universidade Federal do Rio de Janeiro, Caixa Postal
68528, Rio de Janeiro, RJ 21941-972, Brazil}
\author{Marcelo F. Santos}
\affiliation{Instituto de Física, Universidade Federal do Rio de Janeiro, Caixa Postal
68528, Rio de Janeiro, RJ 21941-972, Brazil}
\author{Gabriel H. Aguilar}
\affiliation{Instituto de Física, Universidade Federal do Rio de Janeiro, Caixa Postal
68528, Rio de Janeiro, RJ 21941-972, Brazil}
\author{Adrián A. Budini}
\affiliation{Consejo Nacional de Investigaciones Cient\'ificas y T\'ecnicas (CONICET),
Centro At\'omico Bariloche, Avenida E. Bustillo Km 9.5, (8400) Bariloche,
Argentina, and Universidad Tecnol\'ogica Nacional (UTN-FRBA), Fanny Newbery
111, (8400) Bariloche, Argentina}
\date{\today}

\begin{abstract}
We calculate in an exact way the conditional past-future correlation for the
decay dynamics of a two-level system in a bosonic bath. Different
measurement processes are considered. In contrast to quantum memory
measures based solely on system propagator properties, here memory effects
are related to a convolution structure involving two system propagators and
the environment correlation. This structure allows to detect memory effects
even close to the validity of the Born-Markov approximation. An alternative
operational-based definition of environment-to-system backflow of
information follows from this result. We provide experimental support to our
results by implementing the dynamics and measurements in a photonic
experiment.
\end{abstract}

\maketitle


Decoherence and dissipation are phenomena induced by the unavoidable
coupling of an open quantum system with its environment. When describing
this kind of system dynamics some important approximations are usually
considered. A paradigmatic example is the Born-Markov approximation (BMA),
which considers that the reservoir is not altered significantly due to the
presence of the system. The BMA has been used extensively, providing
excellent agreement with many experiments such as for example in the context
of quantum optics and magnetic resonance.

The high degree of control on individual quantum systems achieved in the
last years leads to the necessity of characterizing dynamics beyond the BMA 
\cite{breuerbook,vega}. In this regime, the environmental degrees of freedom
are affected and depend on the system state. This property gives the
physical ground for a wide class of witnesses and measures of quantum
non-Markovianity \cite{BreuerReview,plenioReview}, where the
system-environment mutual influence is read in terms of an
environment-to-system backflow of information \cite%
{piilo,breuerDecayTLS,dario,sabrina,paris,acin,indu,EnergyBackFLow,Energy,HeatBackFLow}%
. This phenomenon has been studied through physical variables such as energy
and heat \cite{EnergyBackFLow,Energy,HeatBackFLow}, and observed
experimentally in different setups \cite%
{mari, mfs1, mfs2, alejandra,china,breuerDrift,flowExp}.

Consistently, the definitions of the previous quantum memory measures rely
on the system density matrix evolution or propagator, whose properties in
fact encode the memory effects induced by the system-environment coupling.
Nevertheless, even when a quantum master equation is obtained beyond the
BMA, these memory measures may indicate the absence of any non-Markovian
effect. For example, dynamics characterized by positive time-dependent decay
rates are usually classified as Markovian ones~\cite%
{vega,BreuerReview,plenioReview}.

This drawback is circumvented by \textit{operational quantum memory
approaches}, where different consecutive measurements are performed during
the system evolution \cite{modi,pollock,budini,budiniChina}. Both, a
univocal relation between memory effects and departures from BMA, as well as
consistence with the classical definition of non-Markovianity are achieved
with these techniques. Experimental implementation has been recently
performed~\cite{budiniChina}.

In spite of the previous achievements, the understanding of operational
quantum memory witnesses is in its early days. In fact, phenomenon like
environment-to-system backflow of information and similar memory measures
can be completely characterized after knowing the system density matrix
evolution. In contrast, we notice that this information is not sufficient to
characterize operational approaches, where dynamical memory effects that
arise between consecutive measurement processes are not captured by knowing
solely the unperturbed system dynamics. The main goal of this contribution
is to determine which object may take the role of the system propagator when
characterizing operational memory witnesses, which in turn allows to
detecting memory effects close to the validity of the BMA.

In this work, by using a conditional past-future (CPF) correlation method,
we study memory effects in the decay dynamics of a two-level system coupled
to a bosonic environment. Given that this paradigmatic model admits an exact
solution, main differences between operational and non-operational memory
approaches are deduced. 
The CPF correlation is a minimal operational memory witness that is defined
by the correlation between the outcomes of \textquotedblleft
past\textquotedblright\ and \textquotedblleft future\textquotedblright\
system measurement processes when conditioned to a \textquotedblleft
present\textquotedblright\ system state \cite{budini}. We find that, for
different measurement schemes, this witness is proportional to a convolution
term that involves two system propagators and the bath correlation. This
structure only vanishes when approaching the BMA, which elucidates our main
guiding question. An alternative formulation of the phenomenon of
environment-to-system backflow of information, which involves the
measurement processes, follows from this result. We also develop a photonic
setup that implements the system channel dynamics, which provides
experimental support to our findings. Experimental conditions necessary to
achieve resolution close to the Markovian limit are analyzed in detail.

\textit{Microscopic dynamics}: The decay dynamics of a two-level system
induced by a bosonic bath is described by the Hamiltonian \cite{breuerbook}%
\begin{equation}
H_{\mathrm{tot}}=\frac{\omega _{0}}{2}\sigma _{z}+\sum_{k}\omega
_{k}b_{k}^{\dag }b_{k}+\sum_{k}(g_{k}\sigma _{+}b_{k}+g_{k}^{\ast }\sigma
_{-}b_{k}^{\dag }).  \label{Bosonic}
\end{equation}%
Here, $\sigma _{z}$ is the $z$-Pauli matrix, $\sigma _{+}=\left\vert {%
\uparrow }\right\rangle \left\langle {\downarrow }\right\vert $ and $\sigma
_{-}=\left\vert {\downarrow }\right\rangle \left\langle {\uparrow }%
\right\vert $ are the raising and lowering operators of the qubit in the
natural basis $\{\left\vert {\uparrow }\right\rangle ,\left\vert {\downarrow 
}\right\rangle \}.$ The bosonic operators satisfy the relations $%
[b_{k},b_{k}^{\dag }]=1.$

We assume that the total initial wave vector is $|\Psi _{0}\rangle
=(a\left\vert {\uparrow }\right\rangle +b\left\vert {\downarrow }%
\right\rangle )\otimes |0\rangle ,$ where the environment vacuum state is $%
|0\rangle \equiv \prod_{k}|0\rangle _{k}.$ The qubit state $\rho _{t}$ is
obtained by tracing out the environmental degrees of freedom $\rho _{t}=%
\mathrm{Tr}_{e}[|\Psi _{t}\rangle \langle \Psi _{t}|],$ obtaining, in the
interaction picture,%
\begin{equation}
\rho _{t}=\left( 
\begin{array}{cc}
|a|^{2}|G(t)|^{2} & ab^{\ast }G(t) \\ 
a^{\ast }bG^{\ast }(t) & 1-|a|^{2}|G(t)|^{2}%
\end{array}%
\right) .  \label{Rho(t)}
\end{equation}%
The operator $\rho_t$ fulfills the non-Markovian master equation $(d\rho
_{t}/dt)=\frac{-i}{2}\omega (t)[\sigma _{z},\rho _{t}]+\gamma (t)([\sigma
_{-}\rho _{t},\sigma _{+}]+[\sigma _{-},\rho _{t}\sigma _{+}]).$ The
time-dependent decay rate and frequency are defined as $\gamma (t)+i\omega
(t)=-(d/dt)\ln [G(t)].$ The \textquotedblleft wave vector
propagator\textquotedblright\ $G(t)$ obeys the convoluted evolution%
\begin{equation}
\frac{d}{dt}G(t)=-\int_{0}^{t}f(t-t^{\prime })G(t^{\prime })dt^{\prime },
\label{G}
\end{equation}%
where the memory kernel is defined by the bath correlation $f(t)\equiv
\sum_{k}|g_{k}|^{2}\exp [+i(\omega _{0}-\omega _{k})t].$

\textit{Non-operational memory witnesses}: As is well known \cite%
{breuerDecayTLS} for the model (\ref{Bosonic}), standard memory witnesses
such as the trace distance between initial states and departure from
divisibility, coincide. In fact, the dynamics is considered Markovian if
the rate $\gamma (t)$ is positive. Equivalently, this means that $|G(t)|^{2}$
decays monotonically, giving place to a monotonous decay from the upper
level $\left\vert {\uparrow }\right\rangle $ to the lower state $\left\vert {%
\downarrow }\right\rangle .$ Nevertheless, this regime is not necessarily
within the BMA.

\textit{Operational memory witness}: Memory effects defined from a CPF\
correlation \cite{budini} rely on three system measurement processes. They
are performed at successive times $t_{x}<t_{y}<t_{z}$ during the system
dynamics. The CPF measures the correlation between future and past outcomes,
labeled by indexes $z$ and $x$ respectively, when \textit{conditioned} to a
given fixed outcome $y$ at the intermediate present time. Explicitly,%
\begin{equation}
C_{pf}(t,\tau )|_{y}=\sum_{zx}[P(z,x|y)-P(z|y)P(x|y)]O_{z}O_{x},
\label{CPFExplicit}
\end{equation}%
where in general $P(b|a)$ denotes the conditional probability of $b$ given $%
a.$ Furthermore, $t\equiv t_{y}-t_{x}$ and $\tau \equiv t_{z}-t_{y}$ are the
time elapsed between consecutive measurements. $\{O_{z}\}$ and $\{O_{x}\}$
are the (eigen) values of the measured quantum observables.

The CPF correlation intrinsically depends on which measurement processes are
performed at each time. Here, we consider projective measurements performed
in different directions $\hat{n}$\ in the Bloch sphere.\ Thus, $x=\pm 1,$ $%
y=\pm 1,$ $z=\pm 1$, while $O_{z}=z,\ O_{x}=x.$ Considering the initial
state $|\Psi _{0}\rangle $, $t_{x}=0$, and the unitary dynamics associated
to Eq.~(\ref{Bosonic}), we calculate the exact CPF for two different sets of
directions $\hat{n}$ \cite{suppl}. For the directions $\hat{z}$-$\hat{z}$-$%
\hat{z},$ the CPF correlation when conditioned to $y=-1$ reads%
\begin{equation}
C_{pf}(t,\tau )|_{y=-1}\underset{\hat{z}\hat{z}\hat{z}}{=}\left\{ \frac{%
4|a|^{2}|b|^{2}}{[(1-|G(t)|^{2})|a|^{2}+|b|^{2}]^{2}}\right\} |G(t,\tau
)|^{2}.  \label{zzz}
\end{equation}%
Alternatively, by performing successive measurements in the $\hat{x}$-$\hat{z%
}$-$\hat{x}$ directions, for conditional $y=-1,$ it becomes%
\begin{equation}
C_{pf}(t,\tau )|_{y=-1}\underset{\hat{x}\hat{z}\hat{x}}{=}-\left\{ \frac{1-[2%
\mathrm{Re}(ab^{\ast })]^{2}}{1-|G(t)|^{2}/2}\right\} \mathrm{Re}[G(t,\tau
)].  \label{xzx}
\end{equation}%
In the previous two expressions, the function $G(t,\tau )$ is%
\begin{equation}
G(t,\tau )\equiv \int_{0}^{t}dt^{\prime }\int_{0}^{\tau }d\tau ^{\prime
}f(\tau ^{\prime }+t^{\prime })G(t-t^{\prime })G(\tau -\tau ^{\prime }).
\label{GG}
\end{equation}

Apart from normalization factors proportional to the initial system state ($a$ and $b$) and the propagator $%
G(t)$, both Eqs. (\ref{zzz}) and (\ref{xzx}) are proportional to $G(t,\tau )$ \cite{CPFCero}. Thus, in contrast to
previous approaches, instead of $G(t),$ the memory effects here are
determined by this other contribution. It consists in a convolution involving two
system propagators mediated by the environment correlation. It is simple to
check that $G(t,\tau )\rightarrow 0$ when $f(t)$ approaches a delta
function. Consequently, $G(t,\tau )$ measures departures with respect to the
BMA, even close to its validity. Interestingly, this factor has a simple
physical \textit{operational} meaning.

\textit{Backflow of information}: Given that the underlying dynamics admits
an exact treatment, a simple relation between a non-operational backflow of
information \cite{breuerDecayTLS} and an operational one can be established
as follows: Let us consider that the system is at the initial time in the
upper state, a non-monotonous decay of the conditional probability $%
P(\uparrow ,t|\uparrow ,0)=|G(t)|^{2}$ determines the presence of an
environment-to-system backflow of information (non-operational way). In
contrast, under the same initial condition, an operational backflow of
information can be defined by the conditional probability $P(\uparrow
,t+\tau |\downarrow ,t;\uparrow ,0)=|G(t,\tau )|^{2}/[1-|G(t)|^{2}]$ \cite{suppl} which measures the capacity of the environment of reexciting the
system \textit{given} that it has been found in the lower state at an
intermediate time. The previous equality univocally defines the operational
meaning of $G(t,\tau ),$ which in turn guarantees that $P(\uparrow ,t+\tau
|\downarrow ,t;\uparrow ,0)$ only vanishes in the Markovian limit [Eq.~(\ref%
{GG})]. These two clearly different physical scenarios determine the
possibility of detecting or not memory effects close to BMA, which in turn\
may be read as different notions of environment-to-system backflow of
information. These results can be generalized by considering other possible
system states at the initial and final times $[t=0$ and $(t+\tau )$
respectively]. In fact, this degree of freedom leads to different
dependences of the CPF correlation on $G(t,\tau ),$ Eqs.~(\ref{zzz}) and (%
\ref{xzx}).

\textit{Decay channel dynamics:} In order to demonstrate the experimental
feasibility of measuring memory effects close to the BMA, we develop a
photonic platform that simulates the non-Markovian system dynamics. 
The CPF correlation is measured through the sequence $X\rightarrow
U(t)\rightarrow Y\rightarrow U(\tau )\rightarrow Z,$\ where $X,$ $Y,$ and $Z$
are the measurement processes while $U(t)$ and $U(\tau )$ are the unitary
transformation maps associated to the total Hamiltonian (\ref{Bosonic}).
These maps represent the system-environment total changes between
consecutive measurement processes. Although the real environment is composed
of an infinite number of modes, the system reduced dynamical map can be
obtained if the environment is regarded also as a two-level system \cite%
{nielsen}. The map $U(t)$ is defined by the transformations 
\begin{subequations}
\label{eqmap1}
\begin{align}
\left\vert {\downarrow }\right\rangle \otimes \left\vert {0}\right\rangle &
\rightarrow \left\vert {\downarrow }\right\rangle \otimes \left\vert {0}%
\right\rangle , \\
\left\vert {\uparrow }\right\rangle \otimes \left\vert {0}\right\rangle &
\rightarrow \cos (2\theta )\left\vert {\uparrow }\right\rangle \otimes
\left\vert {0}\right\rangle +\sin (2\theta )\left\vert {\downarrow }%
\right\rangle \otimes \left\vert {1}\right\rangle .
\end{align}%
Here, $\left\vert {0}\right\rangle $ and $\left\vert {1}\right\rangle $
represent the bath in its ground state and (first) excited state
respectively. The angle $\theta $ is such that $\cos (2\theta )=G(t).$ Eq. (%
\ref{eqmap1}) is an amplitude damping channel \cite{nielsen}. Given that the
intermediate (second) measurement may leave the system in its ground state
and the bath in an excited state, the channel associated to $U(\tau )$ is
defined as 
\end{subequations}
\begin{subequations}
\label{eqmap2}
\begin{eqnarray}
\left\vert {\downarrow }\right\rangle \otimes |0\rangle &\rightarrow
&\left\vert {\downarrow }\right\rangle \otimes |0\rangle , \\
\left\vert {\uparrow }\right\rangle \otimes |0\rangle &\rightarrow &\cos (%
2\tilde{\theta})\left\vert {\uparrow }\right\rangle \otimes |0\rangle +\sin (%
2\tilde{\theta})\left\vert {\downarrow }\right\rangle \otimes |1\rangle , \\
\left\vert {\downarrow }\right\rangle \otimes |1\rangle &\rightarrow &\sin (%
2\tilde{\theta}^{\prime })\left\vert {\uparrow }\right\rangle \otimes
|0\rangle +\cos (2\tilde{\theta}^{\prime })\left\vert {\downarrow }%
\right\rangle \otimes |1\rangle .\ \ \ \ 
\end{eqnarray}%
This extended damping channel involves one extra initial state, which takes
into account the capacity of the environment of reexciting the system after
it has been found in the ground state. In fact, the angles are given by the
relations $\cos (2\tilde{\theta})=G(\tau ),$ and $\sin (2\tilde{\theta}%
^{\prime })=G(t,\tau )/\sqrt{1-|G(t)|^{2}}$ \cite{suppl}.

The previous maps can be experimentally simulated by encoding the system
states $\{|{\downarrow \rangle },\left\vert {\uparrow }\right\rangle \}$
into polarization of photons $\{|{H\rangle },\left\vert {V}\right\rangle \},$
while the bath states are encoded into the path degree of freedom of
photons. Angles $\{\theta ,\tilde{\theta},\tilde{\theta}^{\prime }\}$ are
chosen as a function of the simulated bath properties \cite{brasil,rio},
where, consistently with Eqs. (\ref{eqmap1}) and (\ref{eqmap2}), a real bath
correlation is assumed $[G(t)\in \mathbb{R}].$

\textit{Experimental setup}: The specific experimental setup is illustrated
in Fig. 1. A continuous-wave (CW) laser, centered at 325 nm, is sent to a
beta-barium-borate (BBO) crystal. Degenerated pairs of photons (wavelength
centered at 650 nm), are produced in the modes signal \textquotedblleft
s\textquotedblright\ and idler \textquotedblleft i\textquotedblright\ via
spontaneous-parametric-down-conversion \cite{Kwiat99}. The photons in mode i
are sent directly to detection as they only herald the presence of photons
in mode s, while the photons in mode s pass through nested interferometers,
which emulate the maps $U(t)$ and $U(\tau )$ \cite{rio}. Projective
measurements are introduced in modules $X,$ $Y,$ $Z.$ The CPF correlation is
extracted by using coincidence counts in each projection set.%
\begin{figure}[tbp]
\includegraphics[bb=19 12 1730 685, width=8.7cm]{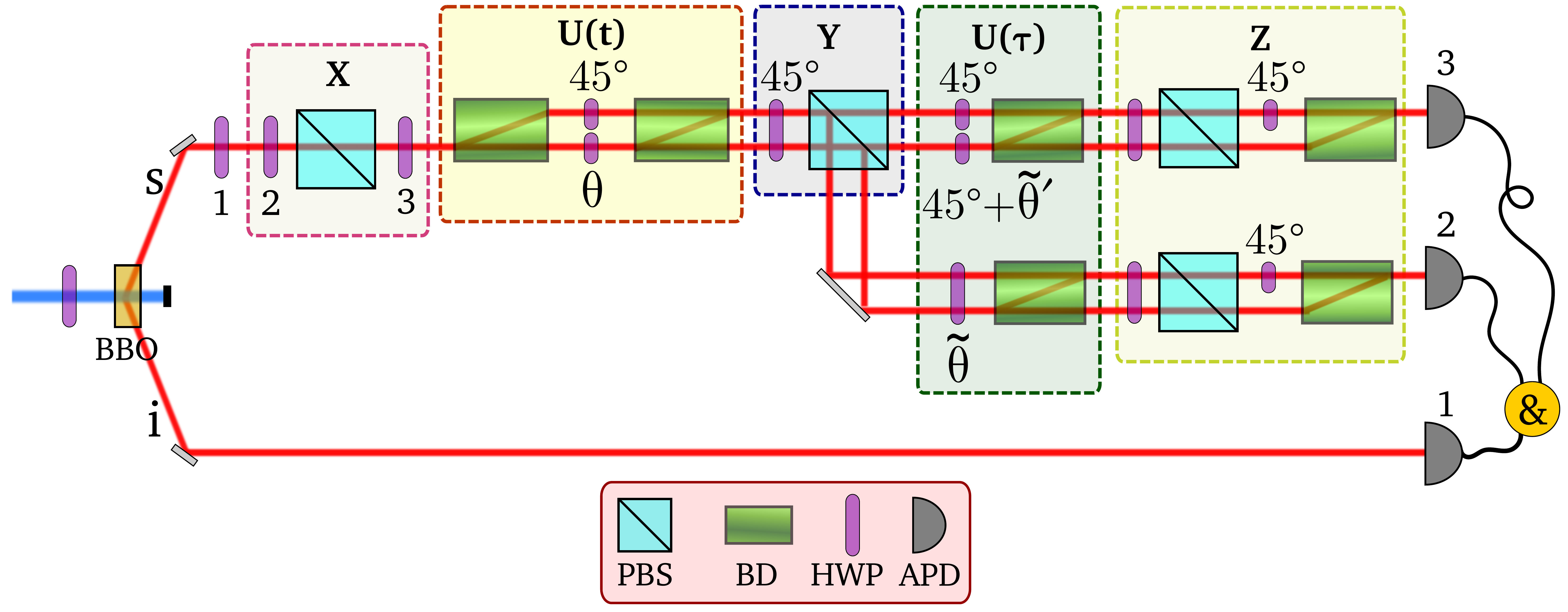}
\caption{Experimental Setup. Modules $X,$ $Y,$ and $Z$ perform the
projective measurements. Modules $U(t)$ and $U(\protect\tau )$ implement the
unitary system-environment maps. From coincidence counting, the avalanche
photon detectors (APD) allow measuring the CPF correlation (see text).}
\end{figure}

Given that the photons created in the BBO crystal are horizontally
polarized, we prepare any initial linear polarization state $(a\left\vert {H}%
\right\rangle +b\left\vert {V}\right\rangle ,\quad a,b\in \mathbb{R})$\
using a half-wave plate (HWP$_{1}$). The past measurement $X$ is performed
using a set of two HWPs and a polarizing beam-splitter (PBS), which
transmits the horizontal polarization and reflects the vertical one. In this
measurement, the angle set in HWP$_{2}$ selects the linear polarization
state mapped to $H$ and hence transmitted by the PBS, while HWP$_{3}$
prepares the projected state from the transmitted horizontal polarization.
After this module, the map $U(t)$ [Eq.~(\ref{eqmap1})] is implemented by
coupling the polarization with the path degrees of freedom. For this, we use
an interferometer composed of two beam-displacers (BD), each transmitting
(deviating) the vertical (horizontal) polarization, and two HWPs, one at
each path mode. HWP$_{\theta }$ rotates the polarization so that photons exit the interferometer in (spatial) mode $|0\rangle $ (upper
path) or in mode $|1\rangle $ (lower path), depending on $\theta $. HWP%
$_{45º}$ simply rotates the photons from $H$ to $V,$ such that all photons
of this mode are mapped to mode $|0\rangle $ at the output of the
interferometer. Posteriorly, measurement $Y$ is performed using a HWP and a
PBS. We restrict ourselves to perform projections in the $\sigma _{z}$ basis.
This is done by fixing a HWP at 45º to correct the polarization state such
that the $H$-polarized photons are transmitted and $V$-polarized ones are
reflected. The map $U(\tau )$ [Eq.~(\ref{eqmap2})], characterized by angles $%
\tilde{\theta}$ and $\tilde{\theta}^{\prime },$ is implemented in a similar
way, noticing that slightly different dynamics take place depending on the
result of the Y measurement ($|\downarrow \rangle $ or $|\uparrow \rangle $,
equivalent here to transmitted or reflected). 
The photons on both paths are coherently combined at the two BD. The final $Z$
measurement is also implemented by two sets of HWP and PBS, one set for the
transmitted light and the other to the reflected light. The last two BDs,
which are just before the detectors Det$_2$ and Det$_3$, are used to trace out the
path degrees of freedom.

From an experimental viewpoint, to condition the probabilities on the result 
$y$ of the intermediate measurement is to consider only the coincidence
counts between Det$_{1}$ and Det$_{3}$ (Det$_{1}$ and Det$_{2}$) for $y=-1$ (%
$y=+1$). Let $N_{z,x}^{(j)}$ denotes the number of coincidences registered
between Det$_{1}$ and Det$_{j}$ when the past and future projective
measurements are set to $x$ and $z$ correspondent eigenvectors,
respectively. The probabilities used to calculate the CPF correlation %
\eqref{CPFExplicit} can be obtained as $P(z,x|y)=N_{z,x}^{(j)}/\left(
\sum_{x^{\prime },z^{\prime }}N_{z^{\prime },x^{\prime }}^{(j)}\right) $,
while $P(z|y)=\sum_{x}P(z,x|y)$ and $P(x|y)=\sum_{z}P(z,x|y)$.

\textit{Detection of memory effects close to the BMA}: The developed
platform allows the simulation of the non-Markovian decay dynamics after knowing
the environment properties. As a concrete example, we consider a bath with a
Lorentzian spectral density, which implies the exponential correlation $%
f(t)=(\gamma /2\tau _{c})\exp [-|t|/\tau _{c}].$ In this case,\ the
propagator (\ref{G}) reads 
\end{subequations}
\begin{equation}
G(t)=e^{-t/2\tau _{c}}\Big{[}\cosh (\frac{t\chi }{2\tau _{c}})+\frac{1}{\chi 
}\sinh (\frac{t\chi }{2\tau _{c}})\Big{]},
\end{equation}%
where $\chi \equiv \sqrt{1-2\gamma \tau _{c}}.$ Furthermore, Eq. (\ref{GG})
becomes%
\begin{equation}
G(t,\tau )=\frac{2\gamma \tau _{c}}{\chi ^{2}}e^{-(t+\tau )/2\tau _{c}}\sinh
(\frac{t\chi }{2\tau _{c}})\sinh (\frac{\tau \chi }{2\tau _{c}}).
\end{equation}%
In the weak coupling limit $\gamma \ll 1/\tau _{c},$ where the correlation
time $\tau _{c}$\ of the bath is the minor time scale of the problem, it
follows that $G(t)\simeq \exp [-\gamma t/2],\ G(t,\tau )\simeq 0,$ which in
turn implies that, independently of the measurement scheme, a Markovian
limit is approached $C_{pf}(t,\tau )|_{y}\simeq 0.$

In Fig. 2 we plot both the theoretical results (full lines) as well as the
experimental ones (symbols) for the CPF correlation at equal times, $%
C_{pf}(t,t)|_{y=-1}.$ Both the $\hat{z}$-$\hat{z}$-$\hat{z}$ [Eq. (\ref{zzz}%
)] and $\hat{x}$-$\hat{z}$-$\hat{x}$ [Eq. (\ref{xzx})] measurement schemes
were measured (upper and lower curves respectively). While for the chosen
bath correlation parameters the propagator $G(t)$ decays in a monotonous
way, detection of memory close to the BMA is confirmed for different bath
correlation times $\tau _{c}.$ An excellent agreement between theory and
experiment is observed. In particular, at time $t=0,$ null values of the CPF
correlation are experimentally observed, meaning that correlation between
the system and environment are negligible at the preparation stage \cite%
{budiniChina}. While the modulus of $C_{pf}(t,t)|_{y=-1}$ depends on the
initial system state, \ we note that it is smaller in the $\hat{z}$-$\hat{%
z}$-$\hat{z}$ scheme when compared with the $\hat{x}$-$\hat{z}$-$\hat{x}$
measurement scheme. In fact, $|G(t,\tau )|^{2}\leq |\mathrm{Re}[G(t,\tau )]|$
[see Eqs.~(\ref{zzz}) and Eq. (\ref{xzx})]. This feature also reflects that
in the former case, in contrast to the last one, the dynamics between
measurements is incoherent.%
\begin{figure}[tbh]
\includegraphics[bb=0 11 370 218,width=8.7cm]{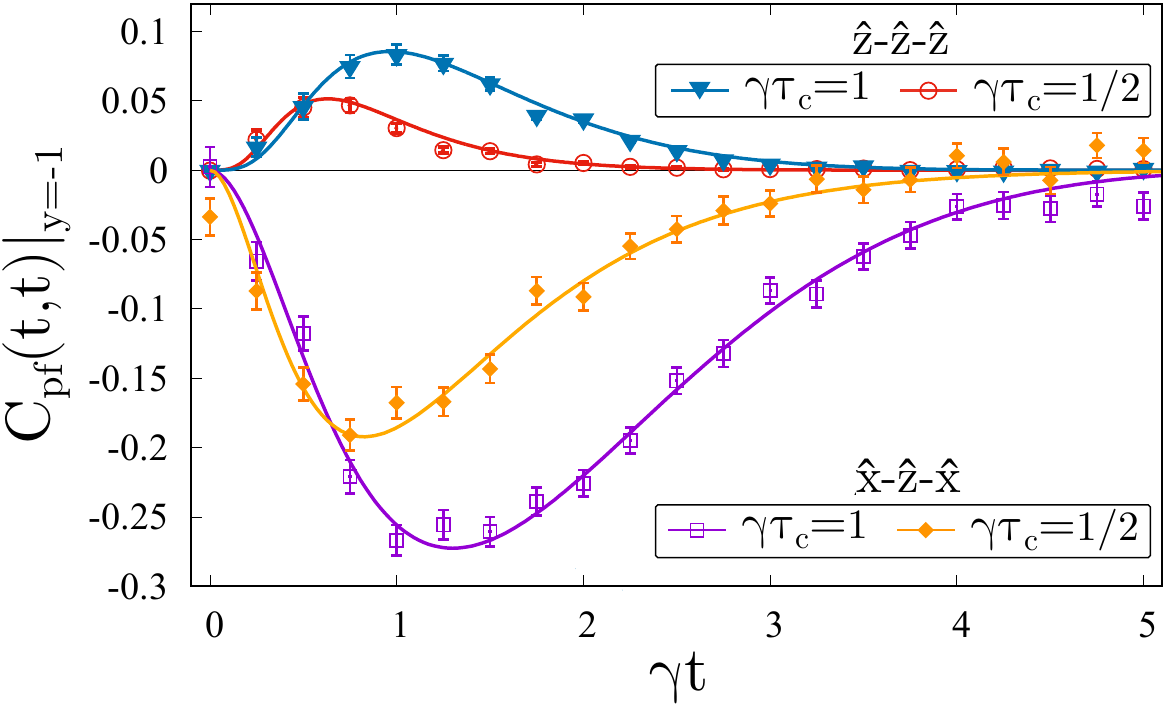}
\caption{CPF correlation for different projective measurements and bath
correlation times. Theoretical results (full lines), experimental results
(symbols). The two upper curves correspond to the $\hat{z}$-$\hat{z}$-$\hat{z%
}$ measurements and the lower ones to $\hat{x}$-$\hat{z}$-$\hat{x}$
measurements. The initial system state is $(\protect\sqrt{p}\left\vert {%
\uparrow }\right\rangle +\protect\sqrt{1-p}\left\vert {\downarrow }%
\right\rangle )$ with $p=0.8$ (upper curves) and $p=1$ (lower curves). From
top to bottom, the bath parameters are $\protect\gamma \protect\tau _{c}=1,$ 
$1/2,$ $1/2,$ $1.$}
\label{fig:CPFexperimental}
\end{figure}

We also used the experimental setup for measuring memory effects even closer
to the BMA, that is, for smaller bath correlation times. Experimental
limitations emerge due to different aspects \cite{suppl}. For instance,
reduced visibility in the interferometers degrades the quality of our
operations, weakening agreement between theory and experiment. The finite
count statistics also become more relevant when approaching the Markovian
limit, as it becomes unclear if a nonnull CPF comes from memory or
fluctuation effects. In spite of these limitations, our experiment
demonstrates the total feasibility of measuring quantum non-Markovian
effects close and beyond the BMA.

\textit{Conclusions}: Detection of quantum non-Markovianity close to the
Born-Markov approximation was characterized through an operational based
memory witness. The CPF correlation was calculated for the decay dynamics of
a two-level system coupled to a bosonic environment. Instead of the
propagator, here the relevant object associated to memory effects consists
in the convolution of two system propagators weighted by the environment
correlation. This structure can be related to an alternative formulation of
the phenomenon of environment-to-system backflow of information, where an
intermediate condition on the system state allows to detects memory effects
even close to the validity of the BMA. A photonic experiment
corroborates the feasibility of detecting quantum memory effects close to
the BMA with excellent agreement with the theory.

These results provide a relevant contribution to the understanding of
operational-based quantum memory witnesses. In particular, our study
elucidates which structure replaces the system propagator when studying
these alternative approaches. The validity of the present conclusions to
arbitrary system-environment dynamics can be established by using
perturbation techniques \cite{bonifacio}.


\textit{Acknowledgments:} T.L.S., M.F.S., S.P.W. and G.H.A. acknowledge financial
support from the Brazilian agencies CNPq (PQ grants 307058/2017-4, 305384/2015-5,
304196/2018-5 and INCT-IQ 465469/2014-0), FAPERJ (PDR10 E-26/202.802/2016, JCN
E-26/202.701/2018, E-26/010.002997/2014, E-26/202.7890/2017), CAPES
(PROCAD2013). A.A.B. acknowledges support from Consejo Nacional de
Investigaciones Científicas y Técnicas (CONICET), Argentina.

\appendix

\section{System-environment unitary evolution}

To calculate the CPF correlation in an exact way, it is necessary to solve
the system-environment dynamics for different initial conditions. The total
Hamiltonian $H_{\mathrm{tot}}=\frac{\omega _{0}}{2}\sigma
_{z}+\sum_{k}\omega _{k}b_{k}^{\dag }b_{k}+\sum_{k}(g_{k}\sigma
_{+}b_{k}+g_{k}^{\ast }\sigma _{-}b_{k}^{\dag }),$ in an interaction
representation with respect to the uncoupled dynamics becomes%
\begin{equation}
H_{\mathrm{tot}}^{I}=\sigma _{+}(t)B(t)+\sigma _{-}(t)B^{\dagger }(t),
\label{Interaction}
\end{equation}%
where $\sigma _{\pm }(t)=\sigma _{\pm }\exp (\pm i\omega _{0}t),$ and $%
B(t)=\sum_{k}g_{k}b_{k}\exp (-i\omega _{k}t).$

\subsection{Evolution in the time interval $(0,t)$}

Let us consider that we can prepare the system-environment in a state%
\begin{equation}
|\Psi _{0}\rangle =(a\left\vert {\uparrow }\right\rangle +b\left\vert {%
\downarrow }\right\rangle )\otimes |0\rangle .  \label{CI}
\end{equation}%
Given that the dynamics of both the system and environment is described by
the Hamiltonian (\ref{Interaction}), the state at time $t$ is written as%
\begin{equation}
|\Psi _{t}\rangle =\Big{[}a(t)\left\vert {\uparrow }\right\rangle
+b(t)\left\vert {\downarrow }\right\rangle +\left\vert {\downarrow }%
\right\rangle \sum_{k}c_{k}(t)b_{k}^{\dag }\Big{]}|0\rangle .
\label{evolved_state_t}
\end{equation}%
From Schrödinger equation, the coefficients evolves as $(d/dt)b(t)=0.$
Therefore, $b(t)=b(0)=b.$ In addition, it follows that%
\begin{eqnarray}
\frac{d}{dt}a(t) &=&-i\sum_{k}g_{k}\exp (+i\phi _{k}t)c_{k}(t), \\
\frac{d}{dt}c_{k}(t) &=&-ig_{k}^{\ast }\exp (-i\phi _{k}t)a(t),
\end{eqnarray}%
where $\phi _{k}\equiv \omega _{0}-\omega _{k}.$ Integrating the last
equation as%
\begin{equation}
c_{k}(t)=c_{k}(0)-ig_{k}^{\ast }\int_{0}^{t}dt^{\prime }\exp (-i\phi
_{k}t^{\prime })a(t^{\prime }),  \label{Ck}
\end{equation}%
the evolution for $a(t)$ becomes%
\begin{equation}
\frac{d}{dt}a(t)=-\int_{0}^{t}f(t-t^{\prime })a(t^{\prime })dt^{\prime
}-ig(t).
\end{equation}%
Here, $f(t)$ defines the bath correlation%
\begin{equation}
f(t)\equiv \sum_{k}|g_{k}|^{2}\exp (+i\phi _{k}t),
\end{equation}%
while the inhomogeneous term is%
\begin{equation}
g(t)\equiv \sum_{k}g_{k}\exp (+i\phi _{k}t)c_{k}(0).
\end{equation}

Defining the Green function $G(t)$ by the evolution%
\begin{equation}
\frac{d}{dt}G(t)=-\int_{0}^{t}f(t-t^{\prime })G(t^{\prime })dt^{\prime },
\label{Propagator}
\end{equation}%
with $G(0)=1,$ the coefficient $a(t)$ can be written as%
\begin{equation}
a(t)=G(t)a(0)-i\int_{0}^{t}G(t-t^{\prime })g(t^{\prime })dt^{\prime }.
\label{ASolution}
\end{equation}

\textit{Initial conditions}:\ Taking the initial conditions%
\begin{equation}
a(0)=1,\ \ \ \ \ \ \ c_{k}(0)=0,  \label{CIFirst}
\end{equation}%
which implies $g(t)=0,$ the coefficients can be expressed as%
\begin{equation}
a(t)=G(t),\ \ \ \ \ \ c_{k}(t)=-ig_{k}^{\ast }\int_{0}^{t}dt^{\prime }\exp
(-i\phi _{k}t^{\prime })G(t^{\prime }).  \label{TimeSolution}
\end{equation}

From Eq. (\ref{evolved_state_t}) if follows that $|a(t)|^{2}$ is the
probability $P(\uparrow ,t|\uparrow ,0)$ of finding the system in the upper
state given that at the initial time $t=0$ it was in the the upper state.
From Eq. (\ref{TimeSolution}) it follows%
\begin{equation}
P(\uparrow ,t|\uparrow ,0)=|G(t)|^{2}.
\end{equation}

\subsection{Evolution in the time interval $(t,t+\protect\tau )$}

To obtain the CPF correlation, the total evolution must also be solved in
the time interval $(t,t+\tau ).$ In this case, the state at time $t$ in Eq. (%
\ref{evolved_state_t}) plays the role of initial state. Letting the
system-environment evolve, the total state can be written as%
\begin{equation}
|\Psi _{\tau }\rangle =\Big{[}\tilde{a}(\tau )\left\vert {\uparrow }%
\right\rangle +\tilde{b}(\tau )\left\vert {\downarrow }\right\rangle
+\left\vert {\downarrow }\right\rangle \sum_{k}\tilde{c}_{k}(\tau
)b_{k}^{\dag }\Big{]}|0\rangle .  \label{evolved_state_tau}
\end{equation}%
In this case, from the Schrödinger equation we get $(d/d\tau )\tilde{b}(\tau
)=0, $ which also implies $\tilde{b}(\tau )=\tilde{b}(0).$ In addition,%
\begin{eqnarray}
\frac{d}{d\tau }\tilde{a}(\tau ) &=&-i\sum_{k}g_{k}\exp [+i\phi _{k}(\tau
+t)]\tilde{c}_{k}(\tau ), \\
\frac{d}{d\tau }\tilde{c}_{k}(\tau ) &=&-ig_{k}^{\ast }\exp [-i\phi
_{k}(\tau +t)]\tilde{a}(\tau ).
\end{eqnarray}%
Integrating the last equation, we have 
\begin{equation}
\tilde{c}_{k}(\tau )=\tilde{c}_{k}(0)-ig_{k}^{\ast }\int_{0}^{\tau }d\tau
^{\prime }\exp [-i\phi _{k}(\tau ^{\prime }+t)]\tilde{a}(\tau ^{\prime }).
\label{ckTilde}
\end{equation}%
As before, replacing this solution in the previous equation we obtain%
\begin{equation}
\frac{d}{d\tau }\tilde{a}(\tau )=-\int_{0}^{\tau }f(\tau -\tau ^{\prime })%
\tilde{a}(\tau ^{\prime })d\tau ^{\prime }-i\tilde{g}(\tau ),
\end{equation}%
where the inhomogeneous term now is%
\begin{equation}
\tilde{g}(\tau )=\sum_{k}g_{k}\exp [+i\phi _{k}(\tau +t)]\tilde{c}_{k}(0).
\label{gTilde}
\end{equation}%
The coefficient $\tilde{a}(\tau )$ can be written in terms of the propagator 
$G(t)$ [Eq. (\ref{Propagator})] as%
\begin{equation}
\tilde{a}(\tau )=G(\tau )\tilde{a}(0)-i\int_{0}^{\tau }G(\tau -\tau ^{\prime
})\tilde{g}(\tau ^{\prime })d\tau ^{\prime }.  \label{SolutionTau}
\end{equation}

To calculate the CPF correlation, we have to consider different initial
conditions.

\textit{First Initial conditions}: When 
\begin{equation}
\tilde{a}(0)=1,\ \ \ \ \ \ \tilde{c}_{k}(0)=0,  \label{CITilde}
\end{equation}%
which implies $\tilde{g}(\tau )=0,$ we have the solution 
\begin{subequations}
\label{SolTilde}
\begin{equation}
\tilde{a}(\tau )=G(\tau ),
\end{equation}%
while from Eq. (\ref{ckTilde}) we get%
\begin{equation}
\tilde{c}_{k}(\tau )=-ig_{k}^{\ast }\int_{0}^{\tau }d\tau ^{\prime }\exp
[-i\phi _{k}(\tau ^{\prime }+t)]G(\tau ^{\prime }).
\end{equation}%
These solutions are equivalent to the previous ones [Eq.~(\ref{TimeSolution}%
)] under the replacement $t\rightarrow \tau .$

\textit{Second initial conditions}: An extra set of initial conditions is
given by 
\end{subequations}
\begin{equation}
\tilde{a}^{\prime }(0)=0,\ \ \ \ \ \ \tilde{c}_{k}^{\prime }(0)=c_{k}(t)/%
\sqrt{1-|G(t)|^{2}},  \label{CITildePrimas}
\end{equation}%
jointly with $c_{k}(0)=0,$ $a(0)=1$ [Eq. (\ref{CIFirst})]. This case
corresponds to finding the system in its ground state after the second
measurement. From Eq.~(\ref{Ck}) we write $c_{k}(t)=-ig_{k}^{\ast
}\int_{0}^{t}dt^{\prime }\exp (-i\phi _{k}t^{\prime })G(t^{\prime }).$ Thus,
Eq.~(\ref{gTilde}) becomes%
\begin{eqnarray}
\tilde{g}(\tau ) &=&-i\frac{\int_{0}^{t}dt^{\prime }\sum_{k}|g_{k}|^{2}\exp
[+i\phi _{k}(\tau +t-t^{\prime })]G(t^{\prime })}{\sqrt{1-|G(t)|^{2}}}, 
\notag \\
&=&\frac{-i}{\sqrt{1-|G(t)|^{2}}}\int_{0}^{t}dt^{\prime }f(\tau +t-t^{\prime
})G(t^{\prime }).
\end{eqnarray}%
From Eq. (\ref{SolutionTau}), $\tilde{a}^{\prime }(\tau )=-i\int_{0}^{\tau
}G(\tau -\tau ^{\prime })\tilde{g}(\tau ^{\prime })d\tau ^{\prime },$
resulting in
\begin{equation}
\tilde{a}^{\prime }(\tau )=\frac{-\int_{0}^{\tau }d\tau ^{\prime
}\int_{0}^{t}dt^{\prime }f(\tau ^{\prime }+t-t^{\prime })G(\tau -\tau
^{\prime })G(t^{\prime })}{\sqrt{1-|G(t)|^{2}}},  \label{Aprevio}
\end{equation}%
which can be rewritten as 
\begin{subequations}
\label{SolTildePrima}
\begin{equation}
\tilde{a}^{\prime }(\tau )=\frac{-G(t,\tau )}{\sqrt{1-|G(t)|^{2}}}.
\end{equation}%
This equation defines the function $G(t,\tau ).$ Moreover, from Eq. (\ref%
{ckTilde}), the other coefficients read 
\begin{eqnarray}
\tilde{c}_{k}^{\prime }(\tau ) &=&\frac{1}{\sqrt{1-|G(t)|^{2}}}\Big{\{}%
c_{k}(t)+ig_{k}^{\ast } \\
&&\times \int_{0}^{\tau }d\tau ^{\prime }\exp [-i\phi _{k}(\tau ^{\prime
}+t)]G(\tau ^{\prime },t)\Big{\}}.  \notag
\end{eqnarray}%
The function $G(t,\tau ),$ after a change of integration variables in Eq. (%
\ref{Aprevio}), can be written as 
\end{subequations}
\begin{equation}
G(t,\tau )=\int_{0}^{t}dt^{\prime }\int_{0}^{\tau }d\tau ^{\prime }f(\tau
^{\prime }+t^{\prime })G(t-t^{\prime })G(\tau -\tau ^{\prime }).
\end{equation}

From Eq. (\ref{evolved_state_tau}) it follows that $|\tilde{a}^{\prime
}(\tau )|^{2}$ is the probability $P(\uparrow ,t+\tau |\downarrow
,t;\uparrow ,0)$ of finding the system in the upper state given that both at
time $t$ is was in the lower state and at the initial time $t=0$ in the
upper state. From Eq. (\ref{SolTildePrima}) it follows%
\begin{equation}
P(\uparrow ,t+\tau |\downarrow ,t;\uparrow ,0)=|G(t,\tau
)|^{2}/[1-|G(t)|^{2}].
\end{equation}

\section{Calculation of the CPF correlation}

Here\ we explicitly calculate the CPF\ correlation defined as%
\begin{equation}
C_{pf}(t,\tau )|_{y}=\langle O_{z}O_{x}\rangle _{y}-\langle O_{z}\rangle
_{y}\langle O_{x}\rangle _{y}.
\end{equation}%
Equivalently, $C_{pf}(t,\tau
)|_{y}=\sum_{zx}O_{z}O_{x}[P(z,x|y)-P(z|y)P(x|y)],$ for different possible
measurement schemes. The conditional values explicitly read%
\begin{equation}
\langle O_{x}\rangle _{y}=\sum_{x=\pm 1}xP(x|y),\ \ \ \ \ \ \ \langle
O_{z}\rangle _{y}=\sum_{z=\pm 1}zP(z|y),  \label{Mean}
\end{equation}%
and%
\begin{equation}
\langle O_{z}O_{x}\rangle _{y}=\sum_{z,x=\pm 1}zxP(z,x|y).  \label{TwoMean}
\end{equation}%
Furthermore, $P(z|y)=\sum_{x=\pm 1}P(z,x|y),$ and $P(x|y)=\sum_{z=\pm
1}P(z,x|y).$ Measurement outcomes are indicated by $x,$ $y,$ and $z,$ while
directions in Bloch sphere are denoted with a hat symbol, $\hat{x},$ $\hat{y}%
,$ and $\hat{z}.$

\subsection{First scheme, measurements \^{z}-\^{z}-\^{z}}

The three measurements necessary to obtain the CPF correlations are
performed in the same $\hat{z}-$direction, with corresponding measurement
projectors $\Pi _{\hat{z}=+1}=\left\vert {\uparrow }\right\rangle
\left\langle {\uparrow }\right\vert $ and $\Pi _{\hat{z}=-1}=\left\vert {%
\downarrow }\right\rangle \left\langle {\downarrow }\right\vert .$ The
initial condition is taken as%
\begin{equation}
|\Psi _{0}\rangle =(a\left\vert {\uparrow }\right\rangle +b\left\vert {%
\downarrow }\right\rangle )\otimes |0\rangle .
\end{equation}

After the first $x$-measurement (measurement in the past), the total state
suffers the transformation $|\Psi _{0}\rangle \rightarrow |\Psi
_{0}^{x}\rangle =\Pi _{\hat{z}=x}|\Psi _{0}\rangle /\sqrt{\langle \Psi
_{0}|\Pi _{\hat{z}=x}|\Psi _{0}\rangle }$ resulting in $(x=\pm 1)$%
\begin{equation}
|\Psi _{0}^{x}\rangle =|x\rangle \otimes |0\rangle ,
\end{equation}%
where we disregarded a global phase contribution. The probability of each
option $P(x)=\langle \Psi _{0}|\Pi _{\hat{z}=x}|\Psi _{0}\rangle ,$ reads%
\begin{equation}
P(x=+1)=|a|^{2},\ \ \ \ \ P(x=-1)=|b|^{2}.
\end{equation}

After the first measurement, the system and environment evolve with the
Hamiltonian dynamics during a time interval $t,$ $|\Psi _{0}^{x}\rangle
\rightarrow |\Psi _{t}^{x}\rangle .$ We get,%
\begin{equation}
\begin{tabular}{cccc}
$x$ & \vline & $|\Psi _{t}^{x}\rangle $ & \vline \\ \hline
$+$ & \vline & $[a(t)\left\vert {\uparrow }\right\rangle +\left\vert {%
\downarrow }\right\rangle \sum_{k}c_{k}(t)b_{k}^{\dag }]|0\rangle $ & \vline
\\ 
$-$ & \vline & $\left\vert {\downarrow }\right\rangle \otimes |0\rangle $ & %
\vline%
\end{tabular}%
,  \label{PsiTime}
\end{equation}%
with $a(0)=1,$ $c_{k}(0)=0$ and normalization $|a(t)|^{2}+%
\sum_{k}|c_{k}(t)|=1.$ Thus, from Eq. (\ref{CIFirst}), these coefficients
are explicitly given by Eq. (\ref{TimeSolution}).

Posteriorly, the second $y$-measurement, correspondent to the present, is
performed. The conditional probability of outcomes $y,$ given the previous
outcomes $x,$ is given by $P(y|x)=\langle \Psi _{t}^{x}|\Pi _{\hat{z}%
=y}|\Psi _{t}^{x}\rangle .$ The joint probability of both outcomes is $%
P(y,x)=P(y|x)P(x).$ The retrodicted probability of past outcomes given the
present ones is $P(x|y)=P(y,x)/P(y),$ where $P(y)=\sum_{x}P(y,x).$ We get%
\begin{equation}
\begin{tabular}{ccccccccc}
$y$ & $x$ & \vline & $P(y|x)$ & \vline & $P(y,x)$ & \vline & $P(x|y)$ & %
\vline \\ \hline
$+$ & $+$ & \vline & $|a(t)|^{2}$ & \vline & $|G(t)|^{2}|a|^{2}$ & \vline & $%
1$ & \vline \\ 
$+$ & $-$ & \vline & $0$ & \vline & $0$ & \vline & $0$ & \vline \\ 
$-$ & $+$ & \vline & $1-|a(t)|^{2}$ & \vline & $(1-|G(t)|^{2})|a|^{2}$ & %
\vline & $\frac{(1-|G(t)|^{2})|a|^{2}}{(1-|G(t)|^{2})|a|^{2}+|b|^{2}}$ & %
\vline \\ 
$-$ & $-$ & \vline & $1$ & \vline & $|b|^{2}$ & \vline & $\frac{|b|^{2}}{%
(1-|G(t)|^{2})|a|^{2}+|b|^{2}}$ & \vline%
\end{tabular}%
.  \label{P(x|y)}
\end{equation}

After the second measurement, the total state suffers the transformation $%
|\Psi _{t}^{x}\rangle \rightarrow |\Psi _{t}^{yx}\rangle =\Pi _{\hat{z}%
=y}|\Psi _{t}^{x}\rangle /\sqrt{\langle \Psi _{t}^{x}|\Pi _{\hat{z}=y}|\Psi
_{t}^{x}\rangle }.$ Posteriorly, starting at time $t,$ $|\Psi
_{t}^{yx}\rangle $ evolves with the total unitary dynamics during a time
interval $\tau ,$ leading to the transformation $|\Psi _{t}^{yx}\rangle
\rightarrow |\Psi _{t+\tau }^{yx}\rangle .$ From Eq. (\ref{PsiTime}) the
states conditioned to the output of each measurement are%
\begin{equation}
\begin{tabular}{ccccccc}
$y$ & $x$ & \vline & $|\Psi _{t}^{yx}\rangle $ & \vline & $|\Psi _{t+\tau
}^{yx}\rangle $ & \vline \\ \hline
$+$ & $+$ & \vline & $\left|{\uparrow}\right\rangle \otimes |0\rangle $ & %
\vline & $[\tilde{a}(\tau )\left|{\uparrow}\right\rangle +\left|{\downarrow}%
\right\rangle \sum_{k}\tilde{c}_{k}(\tau )b_{k}^{\dag }]|0\rangle $ & \vline
\\ 
$+$ & $-$ & \vline & $\nexists $ & \vline & $\nexists $ & \vline \\ 
$-$ & $+$ & \vline & $\left|{\downarrow}\right\rangle \frac{%
\sum_{k}c_{k}(t)b_{k}^{\dag }|0\rangle }{\sqrt{1-|a(t)|^{2}}}$ & \vline & $[%
\tilde{a}^{\prime }(\tau )\left|{\uparrow}\right\rangle +\left|{\downarrow}%
\right\rangle \sum_{k}\tilde{c}_{k}^{\prime }(\tau )b_{k}^{\dag }]|0\rangle $
& \vline \\ 
$-$ & $-$ & \vline & $\left|{\downarrow}\right\rangle \otimes |0\rangle $ & %
\vline & $\left|{\downarrow}\right\rangle \otimes |0\rangle $ & \vline%
\end{tabular}%
.
\end{equation}
The solution form $(y,x)=(+,+)$ comes from Eq.~(\ref{CITilde}) [solutions~(%
\ref{SolTilde})], while for $(y,x)=(-,+)$ follows from Eq.~(\ref%
{CITildePrimas}) [solutions~(\ref{SolTildePrima})].

Finally, the third $z$-measurement is performed (measurement in the future).
The probability $P(z|yx)$ of outcome $z$ given the previous outcomes $y$ and 
$x,$ is given by $P(z|yx)=\langle \Psi _{t+\tau }^{yx}|\Pi _{\hat{z}=z}|\Psi
_{t+\tau }^{yx}\rangle .$ The conditional probability of past and future
event is $P(z,x|y)=P(z|y,x)P(x|y),$ where $P(x|y)$ follows from Eq.~(\ref%
{P(x|y)}). We get 
\begin{equation}
\begin{tabular}{cccccccc}
$z$ & $y$ & $x$ & \vline & $P(z|y,x)$ & \vline & $P(z,x|y)$ & \vline \\ 
\hline
$+$ & $+$ & $+$ & \vline & $|\tilde{a}(\tau )|^{2}$ & \vline & $|G(\tau
)|^{2}$ & \vline \\ 
$+$ & $+$ & $-$ & \vline & $0$ & \vline & $0$ & \vline \\ 
$+$ & $-$ & $+$ & \vline & $|\tilde{a}^{\prime }(\tau )|^{2}$ & \vline & $%
\frac{|G(t,\tau )|^{2}|a|^{2}}{(1-|G(t)|^{2})|a|^{2}+|b|^{2}}$ & \vline \\ 
$+$ & $-$ & $-$ & \vline & $0$ & \vline & $0$ & \vline \\ 
$-$ & $+$ & $+$ & \vline & $1-|\tilde{a}(\tau )|^{2}$ & \vline & $1-|G(\tau
)|^{2}$ & \vline \\ 
$-$ & $+$ & $-$ & \vline & $0$ & \vline & $0$ & \vline \\ 
$-$ & $-$ & $+$ & \vline & $1-|\tilde{a}^{\prime }(\tau )|^{2}$ & \vline & $%
\frac{(1-|G(t,\tau )|^{2}-|G(t)|^{2})|a|^{2}}{(1-|G(t)|^{2})|a|^{2}+|b|^{2}}$
& \vline \\ 
$-$ & $-$ & $-$ & \vline & $1$ & \vline & $\frac{|b|^{2}}{%
(1-|G(t)|^{2})|a|^{2}+|b|^{2}}$ & \vline%
\end{tabular}%
.  \label{P(zx|y)}
\end{equation}%
The conditional probability of the last measurement follows from $%
P(z|y)=\sum_{x}P(z,x|y),$ giving%
\begin{equation}
\begin{tabular}{ccccc}
$z$ & $y$ & \vline & $P(z|y)$ & \vline \\ \hline
$+$ & $+$ & \vline & $|G(\tau )|^{2}$ & \vline \\ 
$+$ & $-$ & \vline & $\frac{|G(t,\tau )|^{2}|a|^{2}}{%
(1-|G(t)|^{2})|a|^{2}+|b|^{2}}$ & \vline \\ 
$-$ & $+$ & \vline & $1-|G(\tau )|^{2}$ & \vline \\ 
$-$ & $-$ & \vline & $\frac{(1-|G(t,\tau )|^{2}-|G(t)|^{2})|a|^{2}+|b|^{2}}{%
(1-|G(t)|^{2})|a|^{2}+|b|^{2}}$ & \vline%
\end{tabular}%
.  \label{P(z|y)}
\end{equation}

From Eqs. (\ref{P(x|y)}) and (\ref{P(z|y)}), the expectation values [Eqs.~(%
\ref{Mean}) and (\ref{TwoMean})] read%
\begin{equation}
\langle O_{x}\rangle _{y=1}=1,\ \ \ \ \ \ \langle O_{z}\rangle
_{y=1}=2|G(\tau )|^{2}-1,
\end{equation}%
while from Eq. (\ref{P(zx|y)}) we get%
\begin{equation}
\langle O_{z}O_{x}\rangle _{y=1}=2|G(\tau )|^{2}-1.
\end{equation}%
Thus, it follows 
\begin{equation}
C_{pf}(t,\tau )|_{y=+1}=0.
\end{equation}%
On the other hand, for $y=-1,$ the averages read 
\begin{equation}
\langle O_{x}\rangle _{y=-1}=\frac{(1-|G(t)|^{2})|a|^{2}-|b|^{2}}{%
(1-|G(t)|^{2})|a|^{2}+|b|^{2}},
\end{equation}%
while%
\begin{equation}
\langle O_{z}\rangle _{y=-1}=\frac{(2|G(t,\tau
)|^{2}+|G(t)|^{2}-1)|a|^{2}-|b|^{2}}{(1-|G(t)|^{2})|a|^{2}+|b|^{2}},
\end{equation}%
and%
\begin{equation}
\langle O_{z}O_{x}\rangle _{y=-1}=\frac{(2|G(t,\tau
)|^{2}+|G(t)|^{2}-1)|a|^{2}+|b|^{2}}{(1-|G(t)|^{2})|a|^{2}+|b|^{2}}.
\end{equation}%
The CPF correlation then is%
\begin{equation}
C_{pf}(t,\tau )|_{y=-1}\underset{\hat{z}\hat{z}\hat{z}}{=}\left\{ \frac{%
4|a|^{2}|b|^{2}}{[(1-|G(t)|^{2})|a|^{2}+|b|^{2}]^{2}}\right\} |G(t,\tau
)|^{2}.  \label{ApCPFZZZ}
\end{equation}

\subsection{Second scheme, \^{x}-\^{z}-\^{x}}

In this scheme, the first and last measurements are performed in $\hat{x}-$%
direction, with measurement projector $\Pi _{\hat{x}+ 1}=|+\rangle \langle
+|,$ and $\Pi _{\hat{x}-1}=|-\rangle \langle-|,$ where $|\pm\rangle =(1/%
\sqrt{2})(\left|{\uparrow}\right\rangle \pm \left|{\downarrow}\right\rangle
).$ The intermediate one is realized in $\hat{z}-$direction, with projector $%
\Pi _{\hat{z}= +1}$ and $\Pi _{\hat{z}= -1}$ defined above. The initial
system-environment state is%
\begin{equation}
|\Psi _{0}\rangle =(a\left|{\uparrow}\right\rangle +b\left|{\downarrow}%
\right\rangle )\otimes |0\rangle .
\end{equation}

After the first $x$-measurement $|\Psi _{0}\rangle \rightarrow |\Psi
_{0}^{x}\rangle =\Pi _{\hat{x}=x}|\Psi _{0}\rangle /\sqrt{\langle \Psi
_{0}|\Pi _{\hat{x}=x}|\Psi _{0}\rangle },$ the bipartite state is%
\begin{equation}
|\Psi _{0}^{x}\rangle =\frac{\left|{\uparrow}\right\rangle +x\left|{%
\downarrow}\right\rangle }{\sqrt{2}}\otimes |0\rangle ,
\end{equation}%
where global phase contributions are disregarded. The probability of each
option $(x=\pm 1)$ $P(x)=\langle \Psi _{0}|\Pi _{\hat{x}=x}|\Psi _{0}\rangle
,$ reads%
\begin{equation}
P(x)=\frac{1}{2}|a+xb|^{2}.
\end{equation}

After the previous step, $|\Psi _{0}^{x}\rangle $ evolves unitarily during a time interval $t,$ $|\Psi _{0}^{x}\rangle \rightarrow
|\Psi _{t}^{x}\rangle .$ Using the initial conditions (\ref{CIFirst}) and
their associated solution (\ref{TimeSolution}), we get%
\begin{equation}
|\Psi _{t}^{x}\rangle =\frac{1}{\sqrt{2}}\Big{[}a(t)\left|{\uparrow}%
\right\rangle +x\left|{\downarrow}\right\rangle +\left|{\downarrow}%
\right\rangle \sum_{k}c_{k}(t)b_{k}^{\dag }\Big{]}|0\rangle ,
\label{PsiAfterXyTime}
\end{equation}%
where $a(0)=1$ and $c_{k}(0)=0.$

Posteriorly, the second $y$-measurement is performed. The conditional
probability for the outcomes is $P(y|x)=\langle \Psi _{t}^{x}|\Pi _{\hat{z}%
=y}|\Psi _{t}^{x}\rangle ,$ which gives%
\begin{equation}
P(+|x)=\frac{|a(t)|^{2}}{2},\ \ \ \ \ \ P(-|x)=1-\frac{|a(t)|^{2}}{2},
\end{equation}%
where we used $|a(t)|^{2}+\sum_{k}|c_{k}(t)|^{2}=1.$ This result indicates
that the random variable $y$ is statistically independent of $x,$ $%
P(y|x)=P(y).$ Thus, the joint probability for the first and second outcomes
is $P(y,x)=P(y|x)P(x)=P(y)P(x).$ The retrodicted probability $%
P(x|y)=P(y,x)/P(y),$ where $P(y)=\sum_{x}P(y,x),$ becomes%
\begin{equation}
P(x|y)=P(x).  \label{Pretro}
\end{equation}

After the second measurement, the state suffers the transformation $|\Psi
_{t}^{x}\rangle \rightarrow |\Psi _{t}^{yx}\rangle =\Pi _{\hat{z}=y}|\Psi
_{t}^{x}\rangle /\sqrt{\langle \Psi _{t}^{x}|\Pi _{\hat{z}=y}|\Psi
_{t}^{x}\rangle }.$ From Eq.~(\ref{PsiAfterXyTime}), for $y=+1$ we get%
\begin{equation}
|\Psi _{t}^{+,x}\rangle =\left|{\uparrow}\right\rangle \otimes |0\rangle ,
\label{CIPlus}
\end{equation}%
while for $y=-1,$%
\begin{equation}
|\Psi _{t}^{-,x}\rangle =\frac{1}{\sqrt{2-|a(t)|^{2}}}\left|{\downarrow}%
\right\rangle \otimes \Big{[}x+\sum_{k}c_{k}(t)b_{k}^{\dag }\Big{]}|0\rangle
.  \label{CIMinus}
\end{equation}

Starting at time $t,$ $|\Psi _{t}^{yx}\rangle $ evolves with the total
unitary dynamics during a time interval $\tau ,$ leading to the
transformation $|\Psi _{t}^{yx}\rangle \rightarrow |\Psi _{t+\tau
}^{yx}\rangle .$ From Eq.~(\ref{CIPlus}) we get%
\begin{equation}
|\Psi _{t+\tau }^{+,x}\rangle =\Big{[}\tilde{a}(\tau )\left|{\uparrow}%
\right\rangle +\left|{\downarrow}\right\rangle \sum_{k}\tilde{c}_{k}(\tau
)b_{k}^{\dag }\Big{]}|0\rangle ,
\end{equation}%
with $\tilde{a}(0)=1,$ $\tilde{c}_{k}(0)=0$ [Eq. (\ref{CITilde})], with $|%
\tilde{a}(\tau )|^{2}+\sum\nolimits_{k}|\tilde{c}_{k}(\tau )|^{2}=1.$ Thus, $%
\tilde{a}(\tau )$ and $\tilde{c}_{k}(\tau )$ are given by Eq.~(\ref{SolTilde}%
). On the other hand, from Eq. (\ref{CIMinus}), it follows that%
\begin{eqnarray}
|\Psi _{t+\tau }^{-,x}\rangle &=&\frac{x\left|{\downarrow}\right\rangle
\otimes |0\rangle }{\sqrt{2-|a(t)|^{2}}}+\sqrt{\frac{1-|a(t)|^{2}}{%
2-|a(t)|^{2}}} \\
&&\times \Big{[}\tilde{a}^{\prime }(\tau )\left|{\uparrow}\right\rangle
+\left|{\downarrow}\right\rangle \sum_{k}\tilde{c}_{k}^{\prime }(\tau
)b_{k}^{\dag }\Big{]}|0\rangle ,  \notag
\end{eqnarray}%
where $\tilde{a}^{\prime }(0)=0$ and $\tilde{c}_{k}^{\prime }(0)=c_{k}(t)/%
\sqrt{1-|a(t)|^{2}}$ [Eq.~(\ref{CITildePrimas})] with $|\tilde{a}^{\prime
}(\tau )|^{2}+\sum\nolimits_{k}|\tilde{c}_{k}^{\prime }(\tau )|^{2}=1.$ In
this case, $\tilde{a}^{\prime }(\tau )$ and $\tilde{c}_{k}^{\prime }(\tau )$
are then given by Eq.~(\ref{SolTildePrima}).

At the final stage, the third $z$-measurement is performed, where the
corresponding conditional probability reads $P(z|yx)=\langle \Psi _{t+\tau
}^{yx}|\Pi _{\hat{x}=z}|\Psi _{t+\tau }^{yx}\rangle .$ From the previous
expressions, we get%
\begin{equation}
P(z|+,x)=\frac{1}{2},
\end{equation}%
while%
\begin{equation}
P(z|-,x)=\frac{1}{2}\Big{[}1-zx\frac{G(t,\tau )+G^{\ast }(t,\tau )}{%
2-|G(t)|^{2}}\Big{]}.
\end{equation}%
The CPF probability $P(z,x|y)=P(z|y,x)P(x|y),$ from the previous two
expressions and Eq. (\ref{Pretro}), reads $(y=+1)$%
\begin{equation}
P(z,x|+)=\frac{1}{2}P(x)=\frac{1}{4}|a+xb|^{2},  \label{cpfPyPlus}
\end{equation}%
while $(y=-1)$%
\begin{equation}
P(z,x|-)=\frac{|a+xb|^{2}}{4}\Big{[}1-zx\frac{G(t,\tau )+G^{\ast }(t,\tau )}{%
2-|G(t)|^{2}}\Big{]}.  \label{cpfPyMinus}
\end{equation}

From Eqs. (\ref{cpfPyPlus}) and (\ref{cpfPyMinus}), the conditional
expectation values [Eqs. (\ref{Mean}) and (\ref{TwoMean})] for $y=+1$ read%
\begin{equation}
\langle O_{x}\rangle _{y=+1}=2\mathrm{Re}(ab^{\ast }),\ \ \ \ \ \ \langle
O_{z}\rangle _{y=+1}=0,
\end{equation}%
and%
\begin{equation}
\langle O_{z}O_{x}\rangle _{y=+1}=0,
\end{equation}%
which implies 
\begin{equation}
C_{pf}(t,\tau )|_{y=+1}=0.
\end{equation}%
On the other hand, for $y=-1,$ the averages read 
\begin{equation}
\langle O_{x}\rangle _{y=-1}=2\mathrm{Re}(ab^{\ast }),
\end{equation}%
while%
\begin{equation}
\langle O_{z}\rangle _{y=-1}=-2\mathrm{Re}(ab^{\ast })\frac{G(t,\tau
)+G^{\ast }(t,\tau )}{2-|G(t)|^{2}}.
\end{equation}%
Furthermore,%
\begin{equation}
\langle O_{z}O_{x}\rangle _{y=-1}=-\frac{G(t,\tau )+G^{\ast }(t,\tau )}{%
2-|G(t)|^{2}}.
\end{equation}%
The CPF correlation then is%
\begin{equation}
C_{pf}(t,\tau )|_{y=-1}\underset{\hat{x}\hat{z}\hat{x}}{=}-\left\{ \frac{1-[2%
\mathrm{Re}(ab^{\ast })]^{2}}{1-|G(t)|^{2}/2}\right\} \mathrm{Re}[G(t,\tau
)].
\end{equation}

For a$\ \hat{y}$-$\hat{z}$-$\hat{y}$ measurements scheme, by performing a
similar calculation, the CPF correlation reads%
\begin{equation}
C_{pf}(t,\tau )|_{y=-1}\underset{\hat{y}\hat{z}\hat{y}}{=}-\left\{ \frac{1-[2%
\mathrm{Im}(ab^{\ast })]^{2}}{1-|G(t)|^{2}/2}\right\} \mathrm{Re}[G(t,\tau
)].  \label{ApCPFXZX}
\end{equation}

\section{Map representation \ of the total unitary dynamics}

For experimental implementation, the system is encoded in the polarization state of single photons, while the bath is effectively implemented through
different spatial modes \cite{brasil,rio}.

The total unitary evolution in first interval $(0,t)$ can be written as the
amplitude damping map
\begin{subequations}
\label{TimeMap}
\begin{eqnarray}
\left\vert {\downarrow }\right\rangle \otimes |0\rangle  &\rightarrow
&\left\vert {\downarrow }\right\rangle \otimes |0\rangle , \\
\left\vert {\uparrow }\right\rangle \otimes |0\rangle  &\rightarrow &\cos
(2\theta )\left\vert {\uparrow }\right\rangle \otimes |0\rangle +\sin (2\theta
)\left\vert {\downarrow }\right\rangle \otimes |1\rangle ,\ \ \ \ 
\end{eqnarray}%
where here $|0\rangle $ and $|1\rangle $ represent spatial modes that
respectively take into account the absence or presence of one excitation in
the environment bosonic modes. Thus, the angle $\theta $ is given by the
relation 
\end{subequations}
\begin{equation}
\cos (2\theta )=a(t)=G(t),
\end{equation}%
where $a(t)$ follows from Eq. (\ref{TimeSolution}).

In the interval $(t,t+\tau )$ the total unitary dynamics realize the
following mapping 
\begin{subequations}
\label{TauMap}
\begin{eqnarray}
\left|{\downarrow}\right\rangle \otimes |0\rangle &\rightarrow &\left|{%
\downarrow}\right\rangle \otimes |0\rangle , \\
\left|{\uparrow}\right\rangle \otimes |0\rangle &\rightarrow &\cos (2\tilde{%
\theta})\left|{\uparrow}\right\rangle \otimes |0\rangle +\sin (2\tilde{\theta}%
)\left|{\downarrow}\right\rangle \otimes |1\rangle , \\
\left|{\downarrow}\right\rangle \otimes |1\rangle &\rightarrow &\sin (2\tilde{%
\theta}^{\prime })\left|{\uparrow}\right\rangle \otimes |0\rangle +\cos (%
2\tilde{\theta}^{\prime })\left|{\downarrow}\right\rangle \otimes |1\rangle
.\ \ \ \ 
\end{eqnarray}%
The angles are given by the relations 
\end{subequations}
\begin{equation}
\cos (2\tilde{\theta})=\tilde{a}(\tau )=G(\tau ),
\end{equation}%
and%
\begin{equation}
\sin (2\tilde{\theta}^{\prime })=\tilde{a}^{\prime }(\tau )=-\frac{G(t,\tau )%
}{\sqrt{1-|G(t)|^{2}}},
\end{equation}%
where $\tilde{a}(\tau )$\ and $\tilde{a}^{\prime }(\tau )$\ follows from
Eqs. (\ref{SolTilde}) and (\ref{SolTildePrima}) respectively.

From the previous mapping, it is possible to rewrite the CPF correlation in
terms of angle variables. From Eq. (\ref{ApCPFZZZ}) we get%
\begin{equation}
C_{pf}|_{y=-1}\underset{\hat{z}\hat{z}\hat{z}}{=}\left\{ \frac{%
4|a|^{2}|b|^{2}}{[\sin ^{2}(2\theta )|a|^{2}+|b|^{2}]^{2}}\right\} \sin
^{2}(2\theta )\sin ^{2}(2\tilde{\theta}^{\prime }),
\end{equation}%
while from Eq. (\ref{ApCPFXZX}) it follows%
\begin{equation}
C_{pf}|_{y=-1}\underset{\hat{x}\hat{z}\hat{x}}{=}\left\{ \frac{1-[2\mathrm{Re%
}(ab^{\ast })]^{2}}{1-\cos ^{2}(2\theta )/2}\right\} \sin (2\theta )\sin (%
2\tilde{\theta}^{\prime }).
\end{equation}%
These two expressions do not depend on angle $\tilde{\theta}.$ In fact, this
angle is relevant when $y=+1,$ where $C_{pf}|_{y=+1}\underset{\hat{z}\hat{z}%
\hat{z}}{=}0$ and $C_{pf}|_{y=+1}\underset{\hat{x}\hat{z}\hat{x}}{=}0.$

The previous expressions for the CPF\ correlation in terms of angle
variables can also be derived from the measurement schemes and by using the
dynamical maps Eqs. (\ref{TimeMap}) and (\ref{TauMap}). For example, the CPF
probability $P(z,x|y)$ for the $\hat{z}$-$\hat{z}$-$\hat{z}$ scheme [compare
with Eq.~(\ref{P(zx|y)})] reads%
\begin{equation}
\begin{tabular}{cccccc}
$z$ & $y$ & $x$ & \vline & $P(z,x|y)$ & \vline \\ \hline
$+$ & $+$ & $+$ & \vline & $\cos ^{2}(2\tilde{\theta})$ & \vline \\ 
$+$ & $+$ & $-$ & \vline & $0$ & \vline \\ 
$+$ & $-$ & $+$ & \vline & $\frac{\sin ^{2}(2\theta )\sin ^{2}(2\tilde{\theta}%
^{\prime })|a|^{2}}{\sin ^{2}(2\theta )|a|^{2}+|b|^{2}}$ & \vline \\ 
$+$ & $-$ & $-$ & \vline & $0$ & \vline \\ 
$-$ & $+$ & $+$ & \vline & $\sin ^{2}(2\tilde{\theta})$ & \vline \\ 
$-$ & $+$ & $-$ & \vline & $0$ & \vline \\ 
$-$ & $-$ & $+$ & \vline & $\frac{\sin ^{2}(2\theta )\cos ^{2}(2\tilde{\theta}%
^{\prime })|a|^{2}}{\sin ^{2}(2\theta )|a|^{2}+|b|^{2}}$ & \vline \\ 
$-$ & $-$ & $-$ & \vline & $\frac{|b|^{2}}{\sin ^{2}(2\theta )|a|^{2}+|b|^{2}}
$ & \vline%
\end{tabular}%
.
\end{equation}%
For the $\hat{x}$-$\hat{z}$-$\hat{x}$ scheme [compare with Eqs.~(\ref%
{cpfPyPlus}) and (\ref{cpfPyMinus})] it can be written as $(y=+1)$%
\begin{equation}
P(z,x|+)=\frac{1}{4}|a+xb|^{2},
\end{equation}%
while $(y=-1)$ 
\begin{equation}
P(z,x|-)=\frac{1}{4}|a+xb|^{2}\Big{[}1+2zx\frac{\sin (2\theta )\sin (2\tilde{%
\theta}^{\prime })}{2-\cos (2\theta )}\Big{]}.
\end{equation}

\section{Robustness of the experimental setup}

\begin{figure}[t]
\includegraphics[bb=13 11 707 400, width=8.7cm]{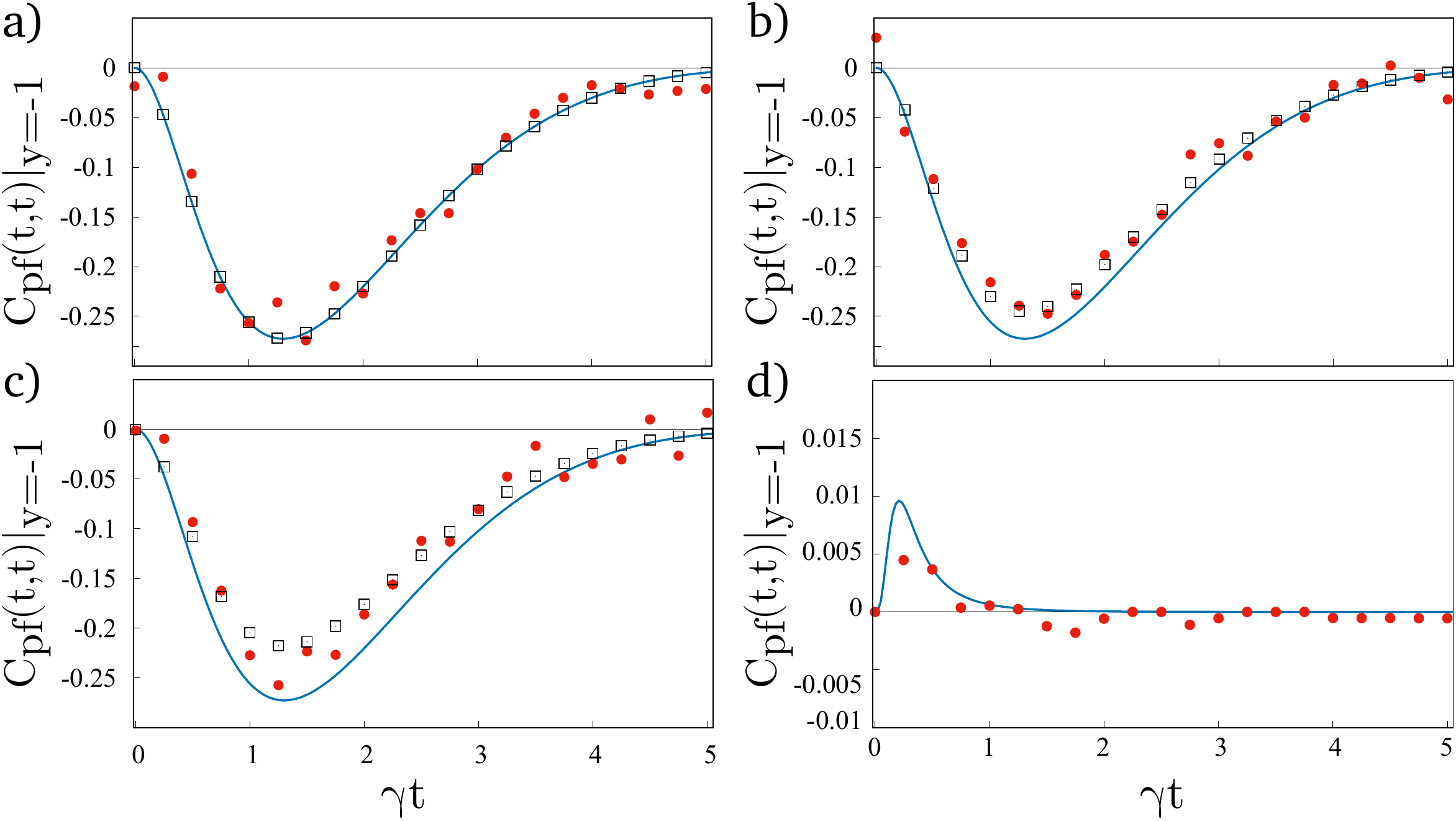}
\caption{CPF correlation as a function of time for simulated data taking
into account different experimental limitations. In all plots the solid blue line represents the ideal theoretical CPF correlation. a), b) and c): V$=1$, V$=0.9$ and V$=0.8$, respectively, for the $\hat{x}$-$%
\hat{z}$-$\hat{x}$ measurement scheme with $\protect\gamma \protect\tau %
_{c}=1,$ and initial state $|\uparrow \rangle $, considering infinite (black squares) or finite (red circles) statistics. In d) the red circles are simulated data considering finite statistics for  the $\hat{z}$-$\hat{z}
$-$\hat{z}$ measurement scheme with initial state $\protect\sqrt{p}%
|\uparrow \rangle +\protect\sqrt{1-p}|\downarrow \rangle $, $p=0.8$. }
\label{fig:CPFsimulado}
\end{figure}

In this section we study the behavior of the CPF correlation in real world
implementations. In particular, we consider two limitations of our
experimental setup, namely the finite counts statistics and the non-unit
visibility of the interferometers. The last one is an issue only for the $%
\hat{x}$-$\hat{z}$-$\hat{x}$ scheme, since the evolution in the $\hat{z}$-$%
\hat{z}$-$\hat{z}$ scheme is incoherent and no interference takes place in
this case. In Fig. \ref{fig:CPFsimulado} we show results of simulations when
these issues are considered. In Fig. \ref{fig:CPFsimulado}a) we show in black
hollow squares the results for the ideal case of visibility V$=1$
and infinite counts. In red circles, we also show results for V$=1$ but
considering finite counts such as the ones we have in the experiment (around 
$10000$ events in total). One can see that the circles are dispersed around
the theoretical prediction, giving rise to values of the CPF correlation up
to 15$\%$ greater than what is expected theoretically. This shows that the
CPF correlation is quite sensitive to statistical fluctuations. In Fig. \ref{fig:CPFsimulado}b) we show results of simulations for V$=0.9$. The results
do not coincide with the theoretical prediction even in the case of infinite
counts (blue hollow squares). Moreover, when imperfect visibility and
finite counts are considered together, experimental values could differ from
theory for more than 25$\%$. When V$=0.8$, results in Fig. \ref%
{fig:CPFsimulado}c), the dispersion of the simulated values is even larger,
obtaining high discrepancy between theory and data. As a consequence, to
restore the agreement between theory and experiment it would be necessary to
introduce dephasing in the theoretical description. By comparing the results of  these simulations and the experimental measurements in Fig. (2) of the main text, we conclude that the main reason for observing some dispersion between experimental data and theory is related to finite counts statistics of our experiment. 

As mentioned in the main text, we find experimental issues when going even
closer to BMA limit ($\tau _{c}\,\rightarrow \,0$). In Fig. \ref%
{fig:CPFsimulado}d) we show the exact value of the CPF correlation (blue
solid curve) and a theoretical simulation including finite statistic effects
(red circles) for $\tau _{c}\gamma =0.1$ in the $\hat{z}$-$\hat{z}$-$\hat{z}$
scheme of measurement. In this case, the values of CPF correlation and its
``experimental'' variations due to fluctuations in the number of counts are
comparable. This alone prevents us to assign a non vanishing correlation to
memory effects instead of considering it as fluctuations.  In this analisis we only took into account finite statistics, making the other experimental issues neglectable.
\begin{figure}[t]
\vspace{10pt} 
\includegraphics[bb=11 12 707 233, width=8.7cm]{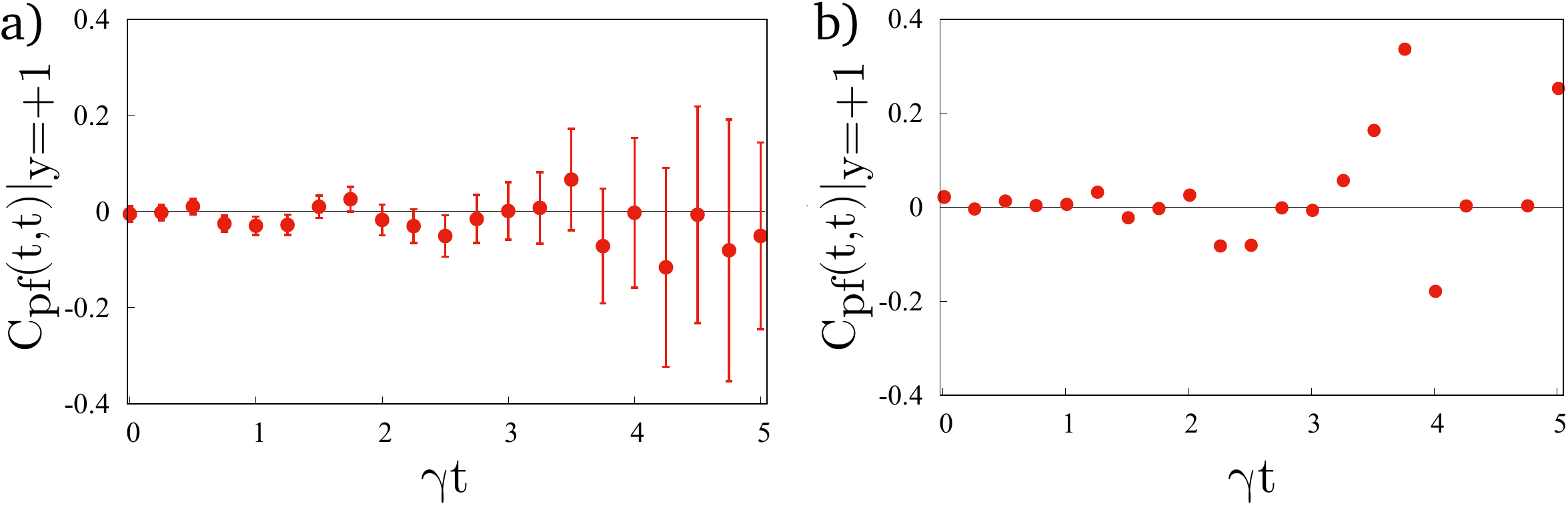}
\caption{CPF correlation for conditional $y=+1$ in the $\hat{x}$-$\hat{z}$-$%
\hat{x}$ scheme. $\protect\gamma \protect\tau _{c}=1,$ and $p=1$ are used in
this case. a) Experimental data and b) simulation assuming Poissonian
fluctuations.}
\label{fig:CPFzero}
\end{figure}

In Fig. \ref{fig:CPFzero}a), we plot the experimental values of CPF
correlation when the outcome of the present (intermediate) measurement is $%
y=+1.$ In this case, the correlation is null within the error bars, in
agreement with what is predicted theoretically. One can see that the error
bars increase substantially while time passes. This is related to the fact
that the system excitation tends to decay to the reservoir, making the
probabilities to find it in an excited state $(y=+1)$ almost null for values
of $\gamma t$ larger that 3. In our setup, this is translated as a reduction
of the number of coincidence counts, causing the probabilities to be much
more sensitive to statistical fluctuations. The fluctuations observed
experimentally are compatible with finite count statistics as shown in Fig. %
\ref{fig:CPFzero}b), where we plot the result of a simulation assuming
Poissonian fluctuations around the ideal theoretical value of the counts.


\begin{thebibliography}{99}
\bibitem{CPFCero} For the conditional $y=+1,$ we get $C_{pf}(t,\tau )|_{y=+1}%
\underset{\hat{z}\hat{z}\hat{z}}{=}C_{pf}(t,\tau )|_{y=+1}\underset{\hat{x}%
\hat{z}\hat{x}}{=}0.$ In fact, when finding the system in the upper state,
the environment must be in the ground state. Therefore, the system evolution
during the first two measurements (interval $t$) is exactly the same than in
the interval\ between the second and third measurements (interval $\tau $).
Consequently, the CPF correlation vanishes~\cite{budini}.
\end{thebibliography}
\end{document}